\documentclass[secnumarabic,floats,amssymb,nobibnotes,nofootinbib,aps,prc,showpacs,tightenlines]{revtex4}
\usepackage[latin1]{inputenc}
\usepackage{pstricks}
\usepackage{amsmath}
\usepackage{float}
\usepackage{color}
\usepackage{bm}
\usepackage{amssymb}
\usepackage[dvips]{graphicx,psfrag,epsfig}

\def \beq  {\begin{equation}}
\def \eeq  {\end{equation}}
\def \ber  {\begin{eqnarray}}
\def \eer  {\end{eqnarray}}

\begin{document}

\title{WMAP Constraints On K-Inflation}

\author{N. Chandrachani Devi$^1$\footnote{chandrachani@gmail.com}, Akhilesh Nautiyal$^2$\footnote{akhilesh@hri.res.in},  Anjan A Sen$^1$\footnote{anjan.ctp@jmi.ac.in}}
\affiliation{$^1$ Centre of Theoretical Physics, Jamia Millia Islamia, New Delhi-110025, India}
\affiliation{$^2$ Harish-Chandra Research Institute, Chhatnag Road,
Jhunsi, Allahabad-211019, India}

\date{\today}

\begin{abstract}
We study the K-Inflation models where the inflaton field has non-canonical kinetic term. In particular, we consider the Dirac-Born-Infeld (DBI) form for the kinetic energy of the inflaton field. We consider quadratic and quartic potentials as well as the potential for the natural inflation. We use a modified version of the MODECODE \cite{Mortonson:2010er} to calculate the power spectrum of the primordial perturbations generated by the inflaton field and subsequently use the WMAP7 results  to constrain the models. Interestingly with DBI type kinetic term, lesser gravity waves are produced as one approaches more towards scale invariance. This is true for all the potentials considered. Unlike the canonical case, this feature, in particular, helps the quartic  ($\lambda\phi^4$) potential  with DBI type kinetic term to be consistent with WMAP data. 
\end{abstract}

\maketitle

\section{Introduction}
In the last two decades, there have been number of breakthroughs in our understanding of the Universe due to some remarkable observational results.  COBE first detected the anisotropy in the cosmic microwave background radiation (CMBR) in 1992 \cite{Smoot:1992td}; subsequently WMAP \cite{Spergel:2003cb,Peiris:2003ff,Spergel:2006hy,Komatsu:2008hk,Komatsu:2010fb}
 measured these anistropies with extraordinary  accuracy  that revolutionized our understanding of the physical Universe (also see the recent results by Atacama Cosmology Telescope (ACT)\cite{Sudeep2011a,Sudeep2011b}).  This has been equally supplemented by the discovery of dark energy by different groups while studying the light curves of different Type-Ia supernovae at different redshifts \cite{sn} as well as the measurement of large scale structure by various galaxy-redshift surveys \cite{lss}. All these observational results strongly support the idea that the early universe had a short period  of accelerated expansion. This epoch of accelerated expansion in the early universe, which is also termed as "inflation", \cite{Guth:1980zm, Linde:1981mu,Starobinsky:1980te}
 not only solves the horizon, flatness and monopole problems in the standard cosmology,
 it is also the simplest and most favored paradigm for the generation of the nearly scalar invariant super-horizon fluctations in the early universe  \cite{Mukhanov:1981xt,Hawking:1982cz,Starobinsky:1982ee,Guth:1985ya},
 which is believed to be the origin of CMBR anisotropy and the large scale structures in our present day Universe.

Simplest model of inflation is usually described by a scalar field (termed as inflaton) rolling over its potential ( also see \cite{Debajyoti:2011} for a different approach for obtaining inflation without scalar fields). The field is minimally coupled to the gravity sector. For a sufficiently flat potential, the potential energy of the field dominates over its kinetic energy which essentially gives rise to negative pressure. This drives the accelerated expansion. This scenario is usually described as ``Slow-Roll Approximation" for inflation \cite{slowroll}.  The primordial perturbations which are responsible for the structures in our universe can be traced back to the quantum fluctuations of this inflaton field in the very early universe. Any given scalar field inflationary scenario can give rise to both the scalar and tensor fluctuations which can be tested against observational data from the CMBR anisotropy measured by WMAP or other experiments. 

With the quality of present data, the primordial fluctation of the inflation field is essentially described by a set of two parameters: the amplitude $A_{s}$ and the spectral index $n_{s}$ of the power specturm of the scalar mode of the density perturbation. As the data from future experiments, such as Planck \cite{planck}, improve, one can add parameters related with primordial gravitational waves or nongaussianity to this set. 

Recently Mortonson et al. \cite{Mortonson:2010er}  have solved the inflationary mode equations numerically without the usual slow-roll approximation. For this, they proposed a new code called "MODECODE" that has been interfaced with the CAMB \cite{Lewis:1999bs}  and COSMOMC \cite{Lewis:2002ah}
 to compute the CMBR anisotropy power spectra and to perform the MCMC analysis for the parameter estimation.  They subsequently used this code to constrain the single scalar field models having canonical kinetic energy term.  They have used the power  law potentials ($V(\phi) = \phi^n$, with $n = 2.3, 1, 2$ and $4$) as well as the natural \cite{Freese:1990rb, Adams:1992bn}  
  and hilltop potentials \cite{hilltop} for their study. The study shows that the current CMB data put stringent constraints on the parameters of a number of well motivated and interesting single field inflationary models. As an example, they showed that the WMAP7 data excludes the quartic potential ($V(\phi) = \lambda \phi^4$) model.

There are large number of attempts to embade inflation in some high energy physics model. One of them is 
k-inflation \cite{kinf}, where the inflation is achieved by the non-standard kinetic terms of the 
inflaton. One of the examples of such a non-canonical kinetic term is the Dirac-Born-Infeld (DBI) form \cite{dbi}. Recently  scalar field with DBI type kinetic term has been widely studied in cosmology for both early and late time accelerated expansion of the Universe \cite{dbiref}. One popular equation of state that arises due to such a kinetic term is so called Chaplygin gas equation of state \cite{gcg} which has many interesting cosmological signatures. 

In the present investigation, we extend the work by Mortonson et al. for scalar fields with DBI type of kinetic term. For this we modify the MODECODE to include the DBI type kinetic term. We study the power law potentials, ($V(\phi) = \phi^n$  with $n =2$ and $4$)  as well as the potential for natural inflation.

\section{Overview of the Calculation}
The action  for the scalar field with a non-canonical kinetic term and minimally coupled to gravity is given by 
\begin{equation}
S=  \int d^4x \sqrt{-g}
\Bigl(\frac{R}{16\pi G} -  V(\phi)\sqrt{1+\eta^2~g^{\mu\nu}
\partial_{\mu}\phi
\partial_{\nu}\phi}\Bigr),
\label{action}
\end{equation}
 where $\eta$ has the dimension of $[length]^2$ and  the field $\phi$  has the dimension of mass.

\subsection{Background evolution}
In a homogenous and isotropic background, the scalar field with non-standard kinetic term has the energy density and presure of form:
\beq
\rho=\frac{V(\phi)}{\sqrt{\left(1-\eta^2 \dot\phi^2\right)}},
\eeq
\beq
P=-V(\phi)(1-\eta^2\dot\phi^2)^{1/2}.
\eeq
Here, dot denotes the differentiation with respect to cosmic time.The evolution of the background  is governed by  the Friedmann's equations :
\beq
H^2=\frac{1}{3M_{pl}^2}\dfrac{V(\phi)}{(1-\eta^2\dot\phi^2)^{1/2}},
\eeq
\beq
\dot{H}=-\dfrac{V(\phi)\eta^2\dot\phi^2}{2M_{pl}^2(1-\eta^2\dot\phi^2)^{1/2}}.
\eeq
The evolution of  the scalar field  is governed by the equation:
\beq
\dfrac{\ddot{\phi}}{(1-\eta^2\dot\phi^2)}+3H\dot\phi+\frac{1}{\eta^2 V(\phi)}\frac{dV(\phi)}{d\phi}=0.
\eeq
In order to solve the equations numerically, we choose number of efolding, $N=\ln a$ as an independent variable and express  the background equations in terms of $N$ as
\ber
H'=-\dfrac{H V(\phi)\eta^2\phi'^2}{2M_{pl}^2(1-H^2\eta^2\phi'^2)^{1/2}}\nonumber \\
\frac{H^2\phi^{\prime\prime}+H H^{\prime}\phi^{\prime}}{\left(1-H^2\eta^2\phi^{\prime2}\right)}+3 H^2\phi^{\prime}+
   \frac{1}{\eta^2 V(\phi)}\frac{dV(\phi)}{d\phi}=0,
\eer
where prime denote the differentiation with respect to N.
\subsection{Perturbation Equations}
The formalism for the perturbations in the K-inflation is given in  \cite{Garriga}.
The evolution of the curvature perturbation $\zeta$ is  
\beq
\ddot{{\zeta}_k}+2\left( \frac{\dot{z}}{z}\right) \dot{\zeta}_k+c_s^2 k^2{\zeta}_k=0,
\eeq
where 
\beq
z= \frac{a\left(\rho+p\right)^{1/2}}{c_s H},
\eeq
 and the effective sound speed 
\beq
c_s^2=\frac{\partial P/\partial\dot{\phi}^2}{\partial\rho/\partial\dot{\phi}^2}=(1-\eta^2\dot\phi^2)=-w.
\eeq

\noindent
Here, dot denotes derivative with respect to conformal time and $k$ is the modulus of the wavevector ${\bf k}$ . The 
curvature perturbation can be expressed in terms of the gauge 
invariant Mukhanov-Sasaki variable 
 $u_k=z{\zeta}$.
The Fourier modes  $u_k$ satisfy the equation
\beq
\ddot{u_k}+\left(c_s^2 k^2-\frac{\ddot{z}}{z}\right)u_k=0.
\label{ukconformal}
\eeq
The equation for $u_k$ in term of e-folding($N$) is
\begin{align}
& u_{k}'' + \left(\frac{H'}{H} + 1 \right)u_{k}' + \left\{\frac{c_s^2 k^2}{a^2 H^2} \right. + 4 + 2\,\frac{H'}{H}+
 \frac{(\phi''+\frac{H'}{H}\phi')}{(1-\eta^2 H^2{\phi'}^2)}\left( \frac{2}{\phi'}+\frac{dV(\phi)/d\phi}{V(\phi)}\right) 
 + 2 \frac{dV(\phi)/d\phi}{\eta^2 H^2V(\phi)\phi'}  \nonumber\\ &
+ \frac{1}{\eta^2 H^2}\left( \frac{1}{V(\phi)}\frac{d^2 V}{d\phi^2}-
\frac{{\left(dV/d\phi\right)}^2}{V(\phi)}\right)u_k =0.
\label{eqscalarper}
\end{align}

Once we solve the $u_{k}$, we can easily find the power spectrum which is  
defined in terms of the two point correlation function as
\beq
{\cal P}_\zeta=\frac{k^3}{2\pi^2}\langle \zeta_k^{\star}\zeta_k^\prime\rangle\delta^3\left(k-k^\prime\right),
\label{powers}
\eeq
and it is related to $u_k$ and $z$ via 
\begin{equation}
{\cal P}_{\zeta}(k)= \frac{k^3}{2\pi^2}\left| \frac{u_k}{z}\right|^2.
\end{equation}
While solving the equations of background and perturbations numerically we use  a modified version of MODECODE \cite{Mortonson:2010er},   to compute the CMBR power spectra, and use COSMOMC \cite{Lewis:2002ah}  to analys the likelihood and perform the parameter estimation.

Tensor pertubations are described by the transverse tracless part of the perturbed metric. Since there is no modification
in tensor perturbations due to non-canonical kinetic term of the scalar field, the Fourier modes of the tensor
 pertubations obey the usual equation:
\beq
\ddot v_k + \left( k^2 - \frac{\ddot a}{a}\right) v_k = 0.
\label{eq:tensmode1}
\eeq
In terms parameter $N$, it becomes
\beq
 v_{k}'' + \left(\frac{H'}{H} + 1 \right)v_{k}' + \left[\frac{k^2}{a^2 H^2} - \left( 2  + \frac{H'}{H} \right) \right]v_k = 0. 
\label{eq:tensmode2}
\eeq
Similar to the scalar power spectrum, the primordial tensor power spectrum is related to $v_{k}$ via
 \beq
{\cal P}^2_t(k) = \frac{4}{\pi^2} \frac{k^3}{M_{pl}^2}\left| \frac{v_k}{a}\right|^2.
\eeq

The scalar and the tensor spectral indices are defined as
\beq
n_{s} = 1 +\left(\frac{{\rm d~ln}{\cal P_{\zeta}}}{{\rm d~ln}k}\right) ~~~{\rm and} ~~~n_{t} = \left(\frac{{\rm d~ln}{\cal P}_t}{{\rm d~ln}k}\right) 
\eeq and the tensor-to-scalar ratio by 
\beq
r = \left(\frac{{\cal P}_t}{{\cal P}_{\zeta}}\right).
\eeq
The scalar spectral index $n_{s}$ and the tensor-to-scalar ratio $ r $ have different values according to the different inflationay models, so they are considered as important obseravable parameters.
Conventionally, a scale invariant power spectrum for scalar perturbations correspond to $ n_s = 1 $ while that for   tensor perturbations correspond to $n_t =0 $.  But for a slow-rolling scalar field, the power spectrum is never an exact scale invariant one. The departure from the scale invariance which is encoded in the spectral index $n_{s}$, is highly constrained by the measurements of the CMBR anisotropy. This puts stringent bound on the parameters of a typical scalar field model. 

\section{Choice of potentials}
We consider the case of chaotic inflation, first introduced by Linde \cite{Linde:1983gd} as well as the natural inflation. 
These models have also been considered by Mortonson et al. \cite{Mortonson:2010er} in case of a scalar field with canonical kinetic energy term. Our motivation is to compare the results obtained by Mortonson et al. with those for similar potentials but with DBI type kinetic term.
First, we consider the potential for chaotic inflation which has the form of 
\beq 
V_(\phi)=\lambda \frac{\phi^n}{n}.
\label{chaotic}
\eeq
Here n=2 corresponds to the qudratic form, $\frac{1}{2}m^2\phi^2$ with $\lambda=m^2$ (m being the mass of the inflaton) and n=4 corresponds to the quartic form $\frac{1}{4}\lambda \phi^4$ ($\lambda$ being dimensionless coupling constant). Both these choices have been extensively studied for the canonical case.  From the earlier studies, it has been found that the mass of the inflaton has to be $10^{12}$GeV and 
$\lambda\le 10^{-12}$ in order to satisfy observational constraints. This value of $\lambda$  for the  quartic case is unnaturally small and it is difficult to embade this in  any reasonable models of particle physics. By considering the non-canonical kinectic term, we expect  different bounds on $\lambda$ and $m^2$ from that of  the standard canonical case.

Next, we consider the model of natural inflation which was 
introduced by Freese et. al. \cite{Freese:1990rb,Adams:1992bn} to explain the small coupling parameter $\lambda$. Here 
inflaton is a Pseudo-Nambu-Goldstone-Boson (PNGB) having potential of the form 
\beq
V(\phi)=\Lambda^4\left(1+\cos\left(\frac{\phi}{f}\right)\right).
\label{natural}
\eeq
Natural inflation in the light of WMAP data has been studied by Savage et. al. \cite{Savage:2006tr} and they showed that the observational constraints on spectral index puts strong bound on the symmetry breaking scale $f\ge 0.7 M_{pl}$.  
 Banks et. al. \cite{Banks:2003sx} have also shown that such a high scale can destabilize the flatness of the potential by large quantum corrections. It was pointed out \cite{Mohanty:2008ab,Mishra:2011vh} that this scale can be reduced to the GUT scale if one considers PNGB coupled with the thermal bath as in warm inflationary models \cite{Berera:1995ie}. So, it  is interesting to see the constraints on such type of potential with a modified kinetic term.
 
 In our analysis, we expressed all the dimensionful quantities in units of the reduced Planck mass.  The priors for the all parameters are given in Table~\ref{tab:priors}. We have set the uniform priors for the common parameters in the models.
The priors for the potential parameters and $\eta$ are
sampled logarithmically as they correspond to some high energy physics scale. We have used a much broader range for the priors 
compared to  the correspoding canonical case (see Table 1 of \cite{Mortonson:2010er}), which is natural 
to do  as we have an extra parameter $\eta$ for kinetic coupling, relaxing the constraints on the potential parameters. We consider the general reheating case where  the constraints on the
total number of e-folds $N_{pivot}$, since the pivot scale left the horizon,  is set as a parameter. So we can study the bound on the value of $N_{pivot}$  in case of non-canonical case using the recent WMAP7 data. The priors for the $N_{pivot}$ has an upper limit given by $N_{irh}$ (similar as \cite{Mortonson:2010er}):
\beq
N_{irh}=55.75-log\left(\frac{10^{16}GeV}{V_{pivot}^{1/4}}\right)+log\left(\frac{V_{pivot}^{1/4}}{V_{end}^{1/4}}\right).
\eeq

\begin{table*}
\caption{Priors on model parameters}
\begin{tabular}{|c|c|c|c|}
\hline 
\hline
Model  & \multicolumn{3}{|c|}{Priors}\\
\hline 
 $n=2$ & $-14<\log m^2<-5$ & $2<\log \eta <10$  & $20<N_{pivot,{\rm}}<70$ \\
 $n=4$ & $-14<\log \lambda<-5$ & $2<\log \eta <10$  & $20<N_{pivot,{\rm}}<70$ \\
 Natural & $-7<\log\Lambda<0$ & $2<\log \eta <14$ & $20<N_{pivot,{\rm}}<70$ \\
          & $-4<\log f<3$ & & \\
           
\hline 
\end{tabular}  
\label{tab:priors}
\end{table*}
%
%
\begin{figure}
\epsfig{file=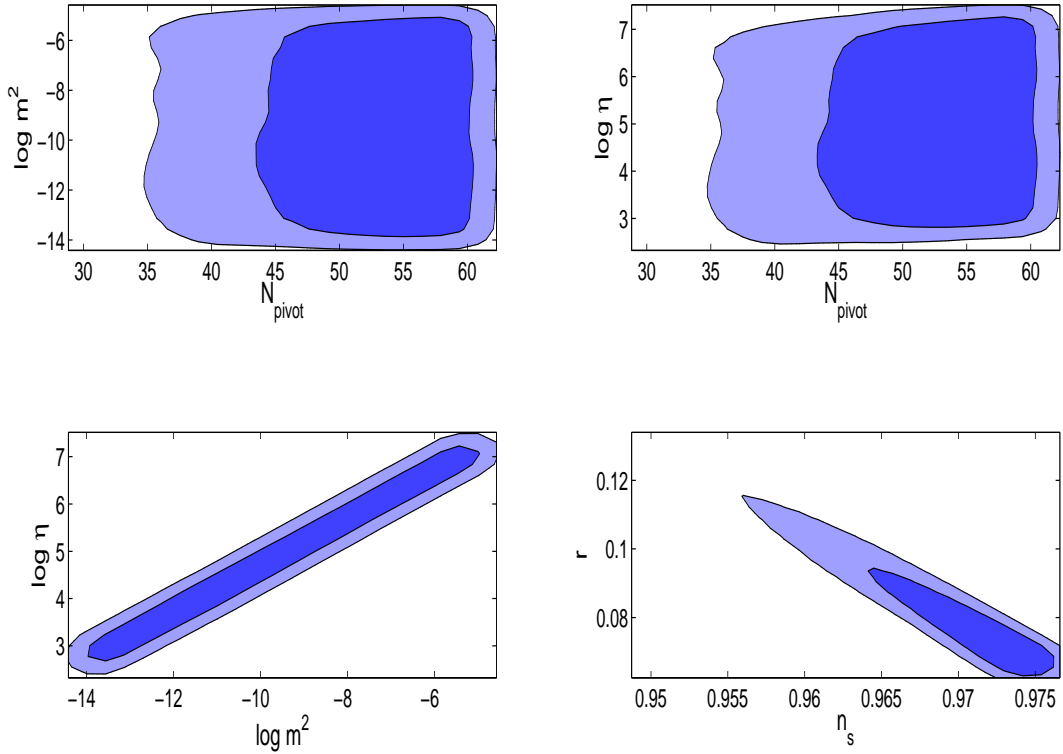, width=17cm,height=12cm}
\caption{Constraints on the quadratic potential ($V(\phi)=m^2\phi^2/2$). Top  Left : $1\sigma$ and $2\sigma$ contours in $\log m^2-N_{pivot}$ phase plane. Top-Right:  $1\sigma$ and $2\sigma$ contours in $\log\eta-N_{pivot}$ phase plane. 
Bottom left: $1\sigma$ and $2\sigma$ contours in  $\log \eta-\log m^2$ phase plane. Bottom right:   $1\sigma$ and $2\sigma$ contours in $n_s-r$ phase plane.}
\label{m2phi2a}
\end{figure}
 \begin{widetext}
\begin{figure}
\epsfig{file=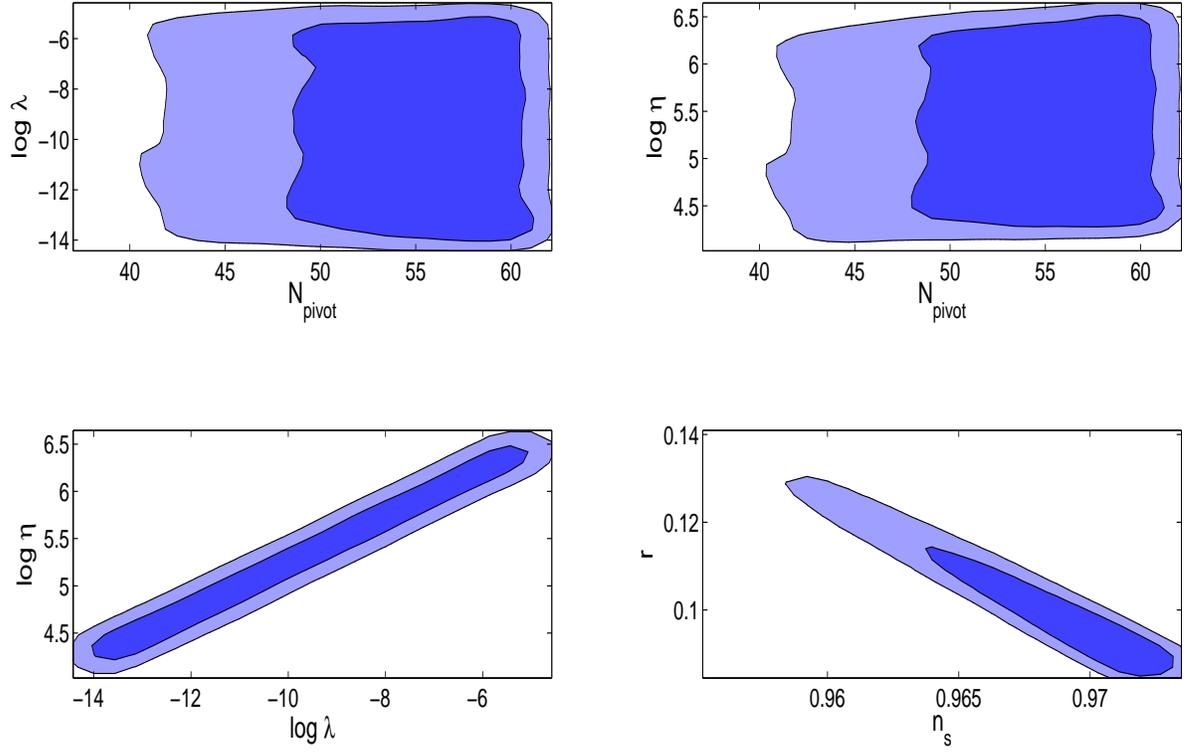, width=19cm,height=12cm}
\caption
{\small\sf  Same as Figure 1, but for  quartic potential, $\lambda \phi^4/4$.}
\label{lphi4a}
\end{figure}
\end{widetext}
\begin{figure}
\epsfig{file=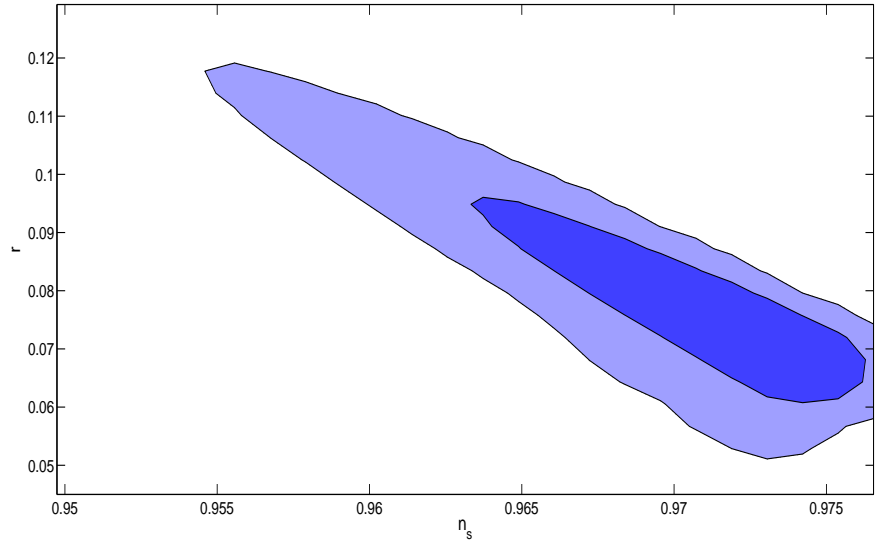, width=14cm,height=8cm}
\caption
{\small\sf  $1\sigma$ and $2\sigma$ contours in $n_s-r$ phase plane for natural inflation  $ V(\phi)=\Lambda^4\left(1+\cos\left(\frac{\phi}{f}\right)\right)$..}
\label{natcon}
\end{figure}

\section{Results}

\begin{figure}
\epsfig{file=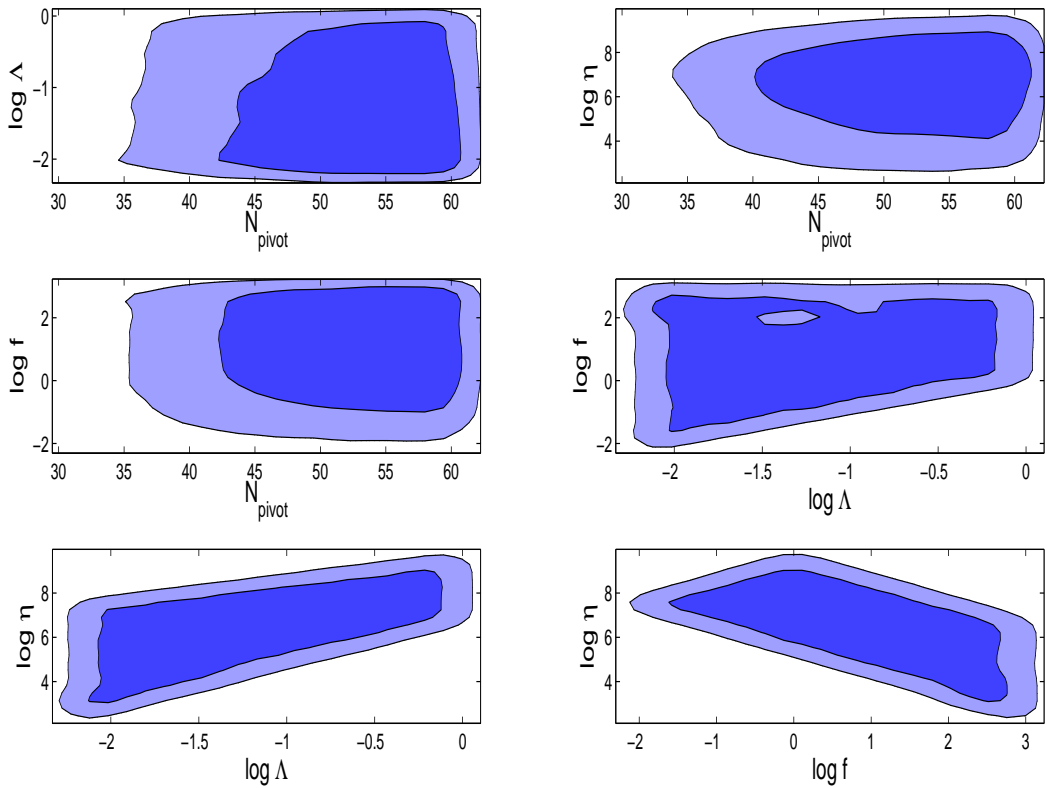, angle=0, width=17cm, height=12cm}
\caption{Same as Figure 1 but for natural inflation, $ V(\phi)=\Lambda^4\left(1+\cos\left(\frac{\phi}{f}\right)\right)$.}
\label{natcon2}
\end{figure}

The results for different potentials are shown in figures ~\ref{m2phi2a}, \ref{lphi4a}, \ref{natcon} and \ref{natcon2}. In Figure 1, we show the constraints on different parameters,  $m^2$, $\eta$, $N_{pivot}$ as well as the allowed region in the $r-n_{s}$ phase plane for the quadratic potential. It is evident from these figures that $N_{pivot}$ is almost uncorrelated to the parameters $m^2$ and $\eta$ in our model, contrary to the canonical case where  $N_{pivot}$ is strongly correlated to the model parameters \cite{Mortonson:2010er}.  Similar behaviour is found for the quartic as well as the natural potential (see figures~\ref{lphi4a}.,~\ref{natcon2}).

Another interesting result is the behaviour of the contours in the  $r$-$n_s$ plane. These contours for all the potentials are  tilted in the opposite direction compared to that of a canonical case. It shows that as the models approach towards scale invariance, one has lesser gravity waves. This makes the models with non canonical kinetic term more probable to satisfy the observations as data always prefer more scale invariance and lesser production of gravity waves.

The best fit values for the parameters are given in 
table~\ref{tab:mlvalues}. 
As evident,  the mean value of $r$ is smaller than 0.1 in all models. The case of $\lambda\phi^4$ potential is particularly interesting. In this case  $r=0.0871$ which is substantially smaller than the value $r=0.27$  obtained for the canonical case \cite{Mortonson:2010er}. With such a small $r$, $\lambda\phi^4$ potential with noncanonical kinetic term is now allowed by the WMAP7 data. Even the Log-Likelihood ($-2\ln\mathcal{L}_{\rm ML}$) value for this model is now drastically reduced and is similar to the values for other potentials. 

Another important result is that the value of $\lambda$ for the quartic potential  can now  be as large as $10^{-7}$ (see table~\ref{tab:mlvalues}) which is a much improved  value compared to the canonical case. But it is still small to be consistent with particle physics.

  \begin{table*}
\caption{Maximum likelihood (ML) values for 
WMAP7 constraints.}
\begin{tabular}{|c|c|c|c|c|c|c|}
\hline 
\hline
Model & Model's parameter&$\log \eta_{\rm ML}$ &$ n_{s,{\rm ML}}$ & $r_{\rm ML}$ &$N_{pivot,{\rm ML}}$& $-2\ln\mathcal{L}_{\rm ML}$ \\
\hline 
 $n=2$ & $\log m^2_{\rm ML}=-9.57$ &5.02 & 0.9745 & 0.0676 & 58.92 & 7475.3579 \\
     
 $n=4$ & $\log\lambda_{\rm ML}=-7.274$ &  5.99 &0.9727 & 0.0871 & 60.9338 & 7475.6259 \\  
        
 Natural & $\log\Lambda_{\rm ML}=-2.19804$  & 5.405 & 0.9733 & 0.0507 & 58.311 & 7475.2481 \\
         
            &  $\log f_{\rm ML}=0.02852$& & & & & \\

\hline 
\end{tabular}  
\label{tab:mlvalues}
\end{table*}

\section{Conclusion}

In brief, we study the inflation driven by the scalar field models with non-canonical kinetic term, more specifically kinetic term of DBI form. To calculate the scalar and tensor power spectrum generated by the quantum fluctuations of these fields, we modify the publicly available MODECODE to incorporate the non-canonical kinetic term. We consider the power law as well as the natural inflation potentials for our study.  There are number of interesting results that emerge from our study. Unlike the canonical case,  parameter $N_{pivot}$, is almost uncorrelated with other model parameters.  
Moreover, the shape of the contours in the $r-n_{s}$ plane for all the potentials show that as one approaches towards scale invariance, lesser gravity waves are produced. This result is completely opposite to that for the canonical case. This helps the $\lambda\phi^4$ potential with a non canonical kinetic term to be allowed by the observational data. There is also a substantial improvement in the Maximum Likelihood value for the $\lambda\phi^4$ potential whereas for other potentials, we get similar number as obtained by Mortonson et al. \cite{Mortonson:2010er} for the canonical case.

In our study, we consider the action where the gravity sector is minimally coupled with the scalar field, having non-standard kinectic term.  This can be easily extended to models where the scalar field is nonminimally coupled with the gravity sector. This will be our aim in future.

\section{Acknowledgement}
The authors are thankful to Hiranya V. Peiris for her suggestions and comments reagarding MODECODE. Numerical computations were performed using the Cluster computing facility in the Harish-Chandra Research Institute,Allahabad, India (http://cluster.hri.res.in/index.html). We acknowledge the use of the MODECODE \cite{Mortonson:2010er} and 
thank the authors to make it publicly available. N.C. Devi acknowledges the financial support provided by the CSIR, Govt. of India. N.C. Devi also acknowledges the Harish-Chandra Research Institute, Allahabad, India for the hospitality and numerical facility provided during various stages of the work. A.A Sen acknoweldges the financial support provided by the SERC, DST, Govt of India through the research grant DST-SR/S2/HEP-043/2009. 



\end{document}